\documentstyle[psfig]{l-aa}
\begin{document}
   \thesaurus{12         
              (11.03.4 A2390;  
               11.05.2;  
               11.19.3;  
               12.03.3;  
               12.07.1)} 

   \title{Redshift survey of gravitational arclets in Abell 2390\thanks{Based 
on observations collected at the Canada-France-Hawaii
Telescope at Mauna Kea, Hawaii, USA}}

   \author{J. B\'ezecourt, G. Soucail}

   \offprints{J. B\'ezecourt}

   \institute{Observatoire Midi-Pyr\'en\'ees, Laboratoire d'Astrophysique, 
    UMR 5572, 14 Avenue E. Belin, F-31400 Toulouse, France }

\date{Received, April 10, 1996, Accepted, June 4, 1996}

\maketitle

\begin{abstract}
In this paper, we present new determinations of redshifts on a sample 
of gravitational 
arclets identified in the cluster of galaxies Abell 2390. The arclets 
candidates were selected from their elongated morphology as seen in 
recent deep HST images of the cluster. The spectra display various 
features, and the redshifts of the background objects range 
from 0.398 to 1.268, most of them being between 0.6 and 0.9 with 
[O\,{\sc ii}] $\lambda$=3727\AA\ being the main emission line encountered.
This distribution is mostly concentrated in two redshifts planes at
$z=0.64$ and $z=0.90$. The existence of two pairs of background galaxies 
at different redshifts, almost superimposed on the same line of sight are 
used to infer some limits on the mass of galaxy-size deflectors. A detailed 
writing of the lensing distorsion by two lens planes at different redshifts 
is developped for this purpose. Finally, 
[O\,{\sc ii}] equivalent widths are used to compute approximate star 
formation rates which appear to be of the same order of magnitude as in 
nearby spiral galaxies. 
The implications on both the lensing properties of the cluster and the 
analysis of the field galaxy evolution are also discussed. 
\end{abstract}

\section{Introduction}
Over the last years, more and more cases of gravitational arcs 
have been discovered bringing important results on 
the mass determination in clusters of galaxies (see Fort and Mellier 
1994 for a review) as well as additional information on the redshift 
measurement of distant galaxies and on their spectral content. 
The use of the gravitational magnification to probe the properties of 
distant background galaxies has already proved to be a powerful method. 
The initial goal in the redshift measurement of arcs was to fix the scaling 
of the mass distribution in the lens modelling (see for example Kneib et 
al., 1993 for the well-known cluster lens Abell 370), at least in the 
central parts of the cluster. But it was also rapidly understood that
clusters could be used as gravitational telescopes to scan and study the
spectral content of a sample of distant field galaxies. In addition, 
Kneib et al. (1994, 1996) have shown that once one arc 
redshift is determined, the knowledge of the lensing potential can be 
used to predict the redshifts of some other lensed sources, from an 
analysis of their shape parameters. From these study they predicted the
redshift distribution of arclets in Abell 370 and Abell 2218. 
This {\sl lensing redshift} method still requires a spectroscopic validation,
at least on several objects in the same cluster, in order to extend it
on fainter and more distant objects. 
The selection of the arclets candidates is much easier with HST deep images
because their distorted morphology is more apparent. 
The gain in spatial resolution has also brought 
spectacular results in the analysis of the size of the spectroscopically
known arcs (Smail et al., 1996).
The redshift-magnitude diagram derived from these redshift estimates 
shows a clear continuity with the existing deep redshift surveys, such 
as those performed by Cowie et al. (1995), Glazebrook et al. (1995) 
or Lilly et al. (1995). These redshift surveys begin to reach the $z \simeq 1$ 
population which is also the case of the arcs redshift survey, but the last 
one corresponds to galaxies typically 1 to 2 magnitudes fainter thanks to the 
gravitational magnification, even if the 
sampling and the selection biases of the objects are quite different.
Very high redshift sources have already been observed thanks to this approach, 
such as the 
source of the giant arc in Cl2244--02 (Mellier et al., 1991), although 
recent observations by Steidel et al. (1996) of a sample of objects 
in the field of high-z quasars has been much more efficient to select 
galaxies at $z\simeq$3--4. 

\begin{figure*}
\psfig{figure=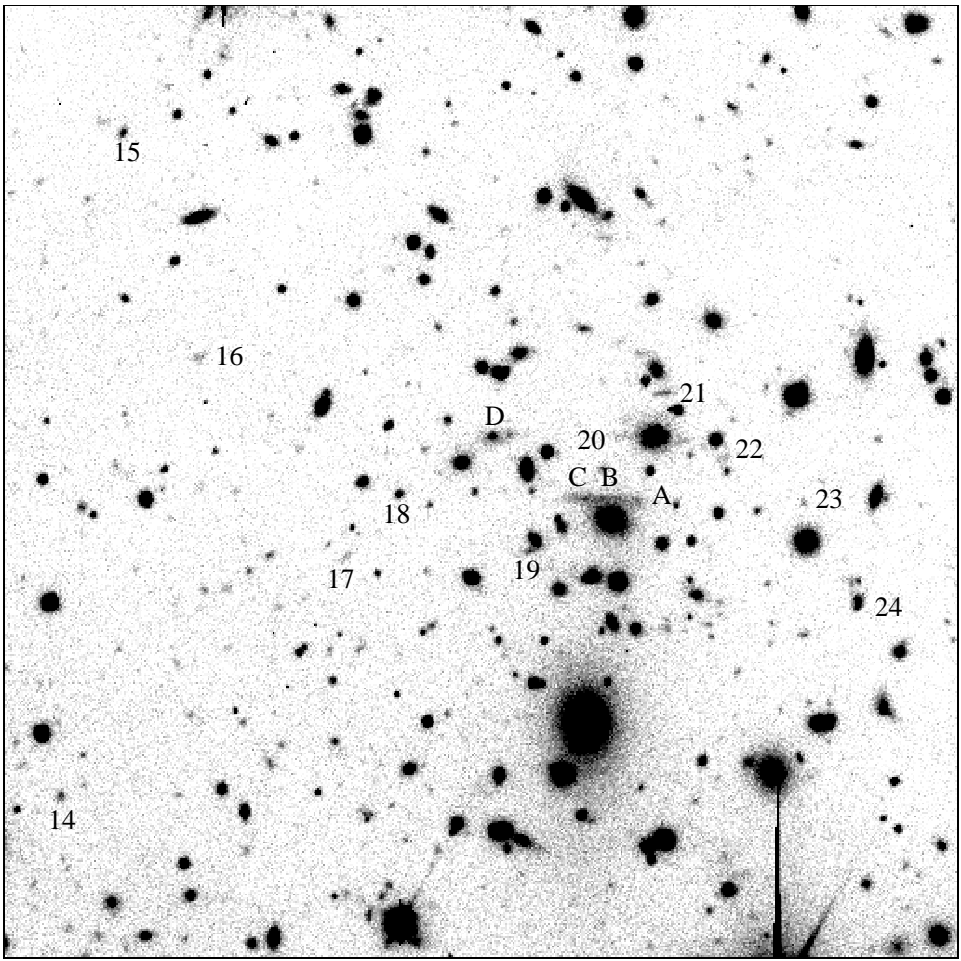,height=14cm}
\caption{R image of the center of Abell 2390, over a 2.7 arcminutes
square field with the 
identification numbers of the objects. Only the central slits are
numbered, with the arclets candidates. 
Although the arclets are faint objects, good seeing conditions at CFHT
allowed their detection in a 10 minutes exposure in R. North is left 
with an angle of $30^\circ$ clockwise.
}
\end{figure*}

In this framework, the recent deep HST images of the cluster Abell 2390 
($z$=0.231) obtained 
by Fort et al. (1996, in preparation) represent an invaluable tool to 
investigate the {\sl lensing redshift} method and its consequences. The 
cluster A2390 is already well known as a cluster-lens which displays 
a ``straight arc'' at a redshift of 0.913 (Pell\'o et al. 1989, 1991) as well 
as many arclets candidates. It was tentatively modelled by Mellier et
al. (1989) and Kassiola et al. (1992), although the proposed models are 
not fully reproducing all the observational constraints. 
Pierre et al. (1996) presented X-ray ROSAT/HRI observations of the cluster  
and an analysis of the matter distribution derived from both the gas 
distribution 
and the lens configuration. The number of arclets identified in the HST 
images is particularly high which makes A2390 one of the best 
lenses to look at the redshift distribution of background objects.
In this paper, we present spectroscopic observations of a sub-sample of the 
arclets candidates. The summary of the observations and the data reduction 
is presented in Section 2, and Section 3 is devoted to the spectral 
analysis of the results. In Section 4 we propose some implications on 
the lensing properties of the field of view and in Section 5, an 
estimate of the star formation rate of the background sources is discussed. 
Finally, Section 6 gives some conclusions and perspectives on the difficult 
problem of redshift determination of faint and distant galaxies.  
Throughout the paper, we consider a Hubble constant of H$_0 = 50$ 
km \,s$^{-1}$\,Mpc$^{-1}$, with q$_0 = 0.5$ and $\Lambda = 0$. 

\section{Observations and data reduction}
Spectroscopic observations were performed on August 24-28, 1995, at
the 3.6m Canada France Hawaii Telescope (CFHT), with the MOS/SIS 
spectrograph (Le F\`evre et al., 1994). The detector was a
2048 $\times$ 2048 Loral 3 CCD with a pixel size of 0.314\arcsec\ and 
an image field of 10\arcmin $\times$ 10\arcmin . Low dispersion 
spectroscopy was obtained with the V150 grism providing a
wavelength dispersion of 7.3\AA\ per pixel. 
One mask was designed with 36 slitlets (1.2\arcsec $\times$
10\arcsec ) and 7 exposures of 1 hour each were obtained in good weather 
conditions (average seeing of 0.9\arcsec ). 

Arclet candidates were essentially selected from the deep
HST/WFPC2 images of the cluster. No a priori color
selection was introduced, and only the shape and the orientation were
taken into account, favoring the highest magnified sources. We imposed a 
selection criterium so that the arclets candidates must have an axis 
ratio $a/b$ greater than 3.5 (either in V or I) and are oriented 
roughly tangentially with respect to the center of the cluster. An 
additional check by eye allowed to eliminate a few faint and spurious 
objects, finally giving a sample of 35 candidates.
Most of them are located in the region of the
giant straight arc, where the lensing power of the cluster is the strongest.
An additional cut in magnitude occured directly on a 10minutes R exposure 
from which the selection was done, as we required to detect the objects.  
From the constraints of the mask preparation and the slit selection, 
only a sub-sample was selected, and the rest of the mask was filled with 
cluster members candidates, in order to increase the redshift data in 
A2390. The selected objects are displayed in Figure 1. 

A mask was punched with 36 slits, 19 corresponding to cluster
members, 3 to M stars and 2 are unidentified. 12 spectra correspond to
objects at higher $z$ than the cluster redshift but 3 of them are ranging 
from 0.3 to 0.4 and located too far away from the center to be significantly 
lensed, and will not be considered in this paper. We will focus only 
on the 11 most central slits, which correspond to objects in the field
of view of the WFPC2 image (namely objects \#15 to \#24) plus object 
\#14. 

Data reduction was performed with standard IRAF routines for bias 
removing and flat-fielding, and the 
Multired package that is especially devoted to multislit spectroscopy.
From the 7 individual exposures, a median averaged sky-subtracted spectrum 
was obtained for each slitlet, allowing an efficient removal of cosmic rays 
and other defects. The extracted spectra were flux calibrated with the 
standard star BD253941 and the shape of the continuum can be considered 
reliable from 4000\AA\ to 8000\AA. Finally, a boxcar smoothing with a window of 
20 \AA\ (3 pixels) was applied to increase the S/N on the faintest spectra. 

\section{Redshift and spectroscopic content of the arclets}
New redshift determinations are based essentially on the identification of 
the [O\,{\sc ii}] emission line $\lambda$3727\AA\ (Figure 2), 
although for some arclets, a more detailed analysis of the spectral content 
is possible (\#21, \#24). The results are presented in Table 1 (results 
concerning the giant arc have been added from Pell\'o et al., 1991).
Over the 11 spectra extracted from the central slits, 
we obtain 12 reliable redshift determinations: 3 objects belong to the cluster, 
and 9 are at larger redshift, $z$ ranging from 0.398 to 1.268. Only 
one spectrum 
(\#19) does not present significant features and has no redshift 
identification, 
giving a success rate of 90\%\ for the redshift measurement of the background  
arclets. This is important to note, as we can exclude any strong 
bias towards the high redshift tail of galaxies,  regardless
the fact that between $z\simeq$ 1.4 and $z\simeq$ 2.2, no strong emission
line falls in the optical band for ``normal'' galaxies.

Although the statistics of the background objects is small, two comments 
can be done here on the redshift distribution of the arclets. First, our 
results confirm the trend observed in the whole sample of cluster lenses, 
for which most sources are between $z \simeq 0.6$ and $z \simeq 1$ 
(Fort and Mellier 1994) and only a few of them are at redshift larger than 
1 (namely \#16$\_$1 and \#16$\_$2 in our sample). 
Second, it seems that two redshift planes or windows may be privileged 
behind Abell 2390.
The first one is at $z\simeq$ 0.64 and displays at least 4 identified
objects: arclets \#20, \#21 and \#24$\_$1, plus one galaxy from Le Borgne
et al., 1991). The other one is at $\simeq$ 0.90 and corresponds to the
redshift of the giant straight arc. It contains arclets \#14 and \#17,
and the two sources of the giant arc (we assume that object A is
different from B--C, as suggested from the HST image) as well as object
D from Pell\'o et al. (1991). The two planes are quite significant as
they are spatially coherent: in both cases, all the identified object
fall inside a radius smaller than 1\arcmin .
We then suspect that two groups or clusters of galaxies are located at 
these redshifts, although a better statistics of arclets redshifts is 
required to reinforce this result. Future observations are planned to 
complete this redshift survey. An alternative strategy will be to study 
the multi-color distribution of objects and the implied photometric 
redshift following the method proposed by Pell\'o et al. (1996), but 
this requires deep photometry of the field over a large wavelength range, 
from B to near-infrared colors (J and K'). 

\begin{figure*}
\vskip -11cm
\hbox{
\hskip -1cm
\vbox{
\psfig{figure=fig2a.ps,width=6cm}
\psfig{figure=fig2b.ps,width=6cm}
\psfig{figure=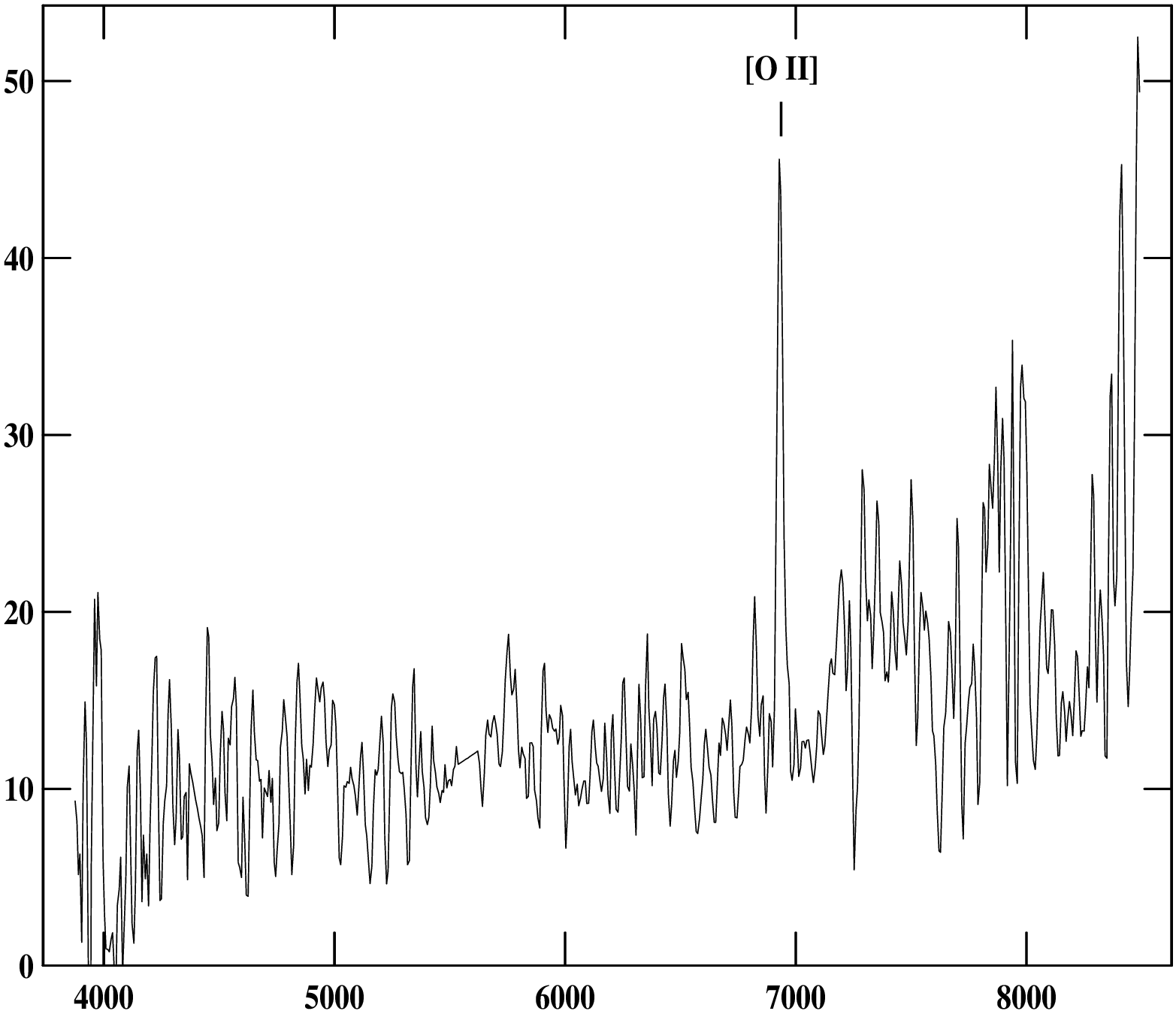,width=6cm}
\psfig{figure=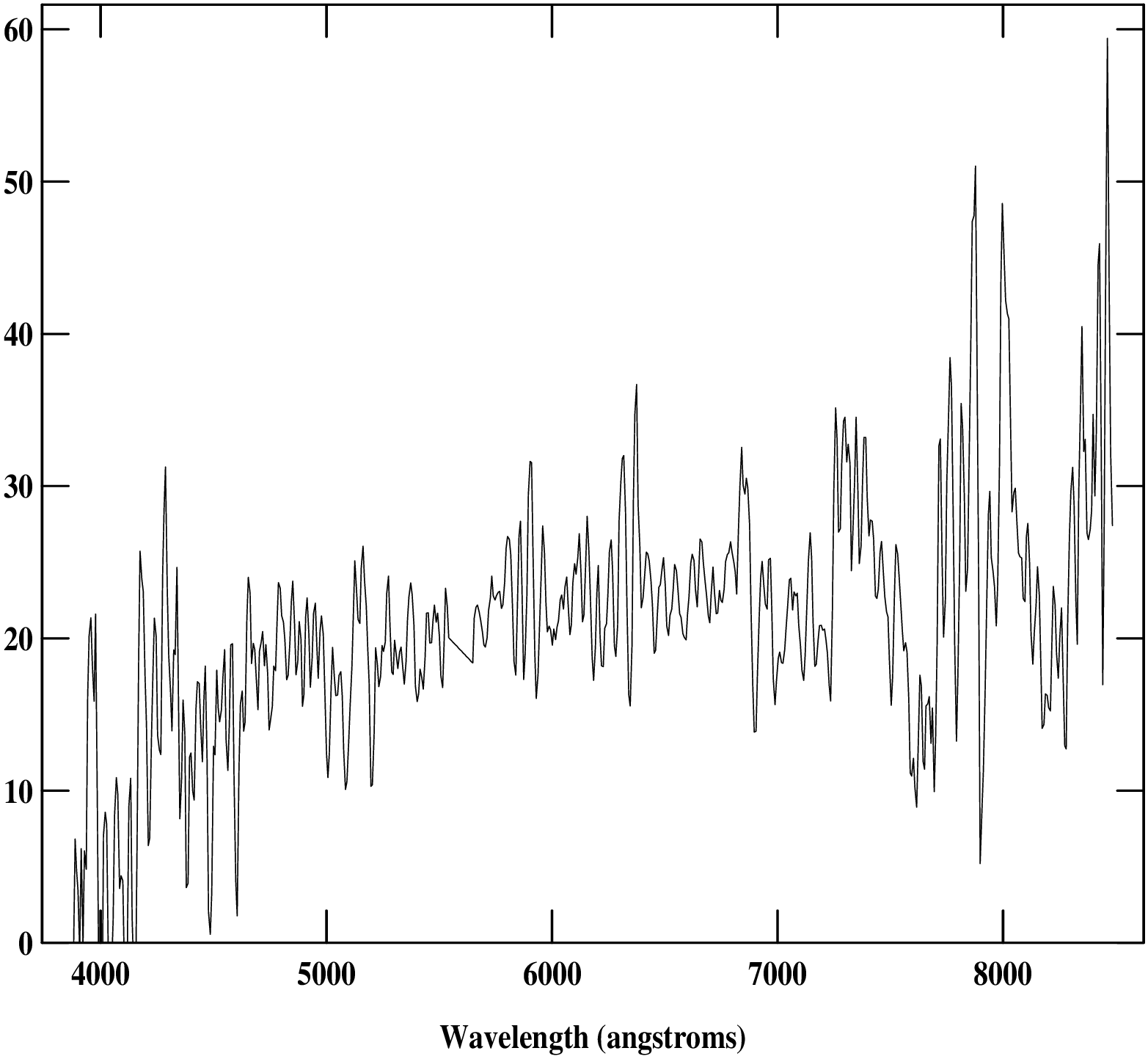,width=6cm}}
\hskip 3.1cm
\vbox{
\psfig{figure=fig2e.ps,width=6cm}
\psfig{figure=fig2f.ps,width=6cm}
\psfig{figure=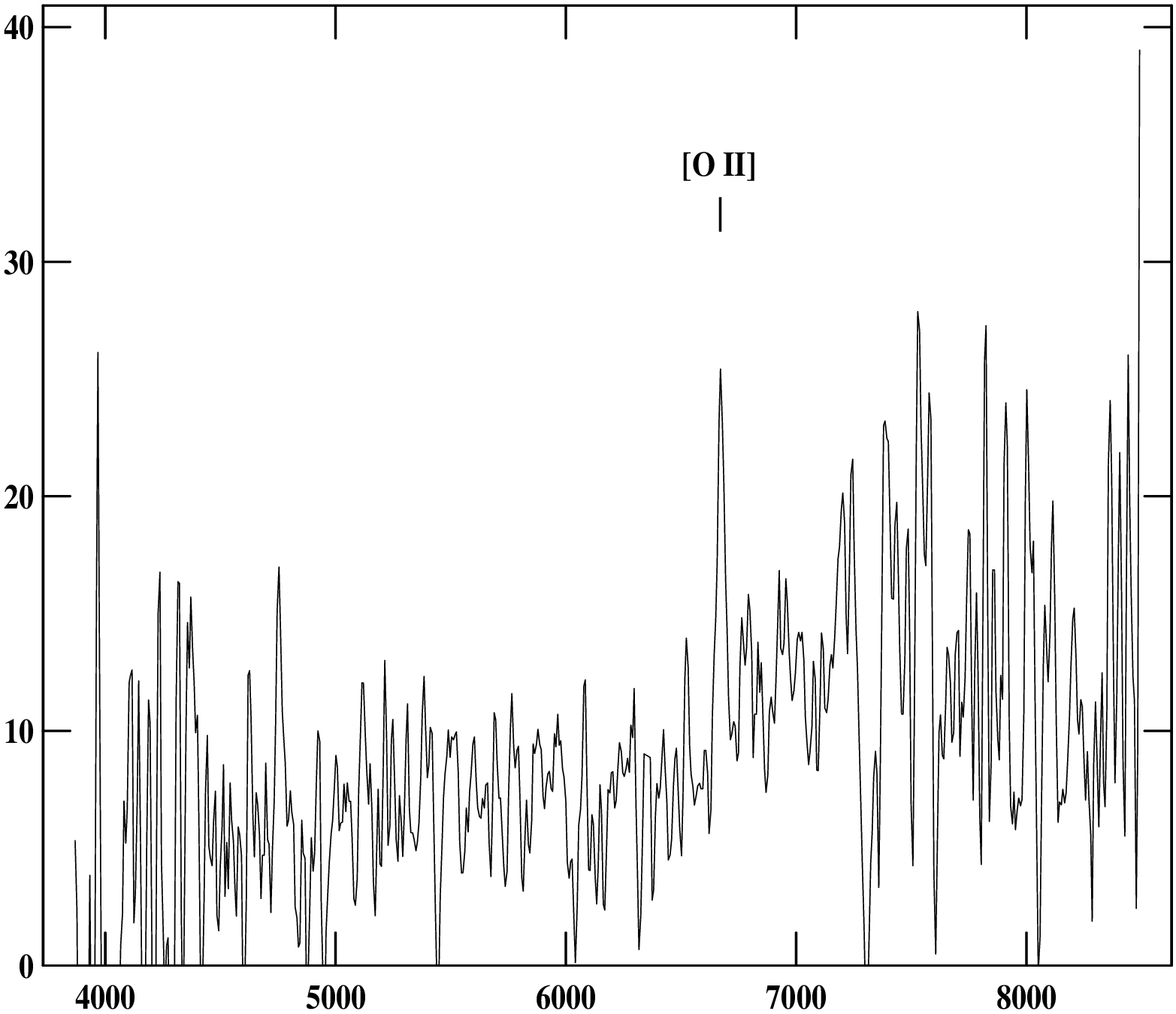,width=6cm}
\psfig{figure=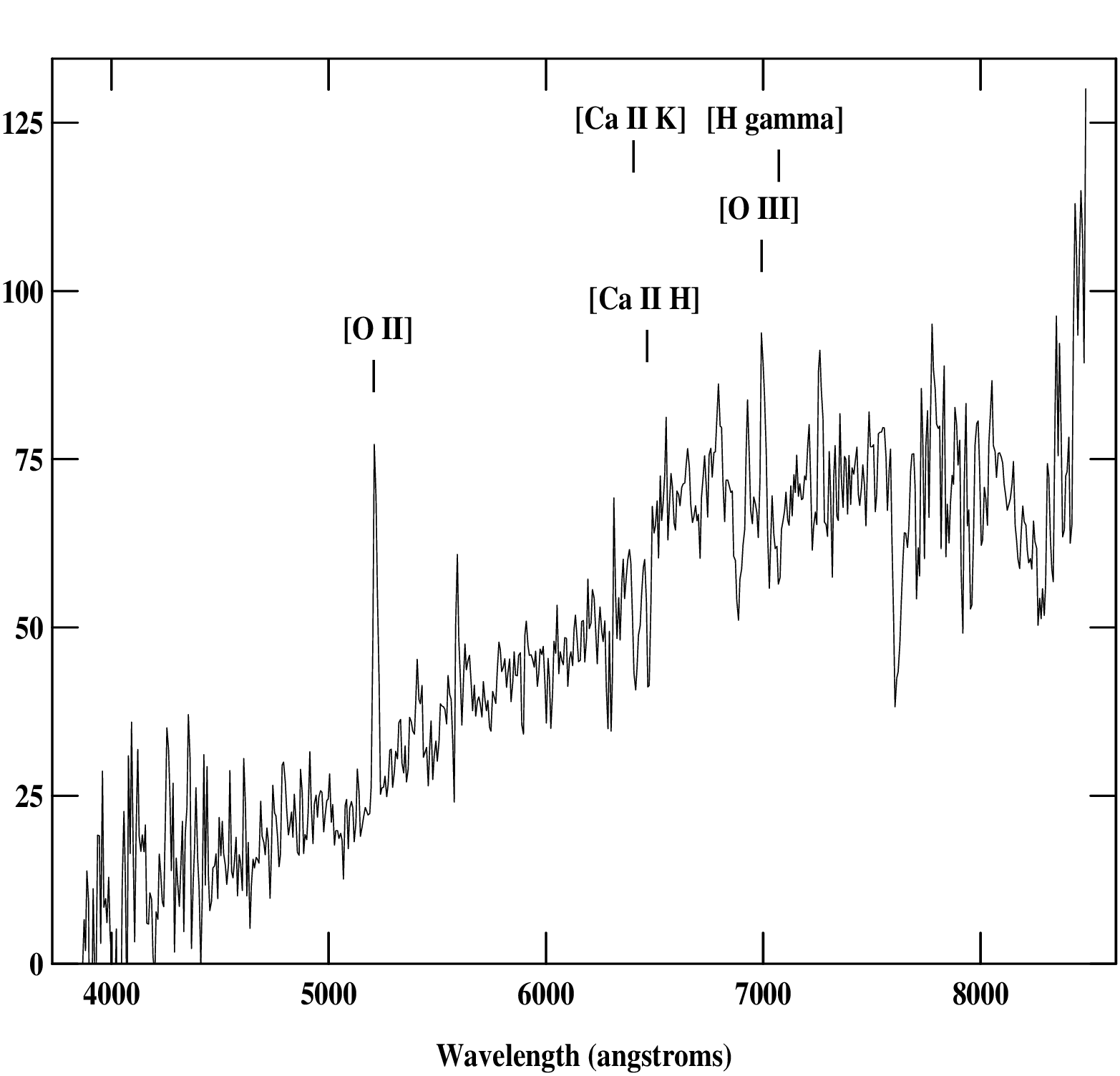,width=6cm}}}
\vskip -20.5cm
\hskip 1cm
\#14
\hskip 8.4cm \#20
\vskip 5.3cm
\hskip 1cm
\#16
\hskip 8.4cm \#21
\vskip 5.3cm
\hskip 1cm
\#17
\hskip 8.4cm \#22 
\vskip 5.3cm
\hskip 1cm
\#19
\hskip 8.4cm \#24
\vskip 4.5cm
\caption{Spectrum of the arclets \#14, \#16, \#17, \#19, \#20, \#21,
\#22 and \#24.
Ordinate is $F_{\lambda}$ in arbitrary units. Spectra have been smoothed
with a window of 20 \AA\ (see text for more details). Identified
features are marked on each spectrum. }
\end{figure*}

\begin{table*}
\caption[]{Spectroscopic properties of arclet candidates in A2390:
The first column is the slitlet number except for the first three
objects taken from Pell\'o et al. (1991).
$W_\lambda$ is the rest frame
equivalent width of [O\,{\sc ii}], B is the magnitude measured on CFHT
images
and M is the gravitational magnification of
the arclet computed from the lensing model of Pierre et al. (1996). The
estimated star formation rate is indicated in the last column (see
Section 5).}
\begin{flushleft}
\begin{tabular}{ccccccc}
\hline\noalign{\smallskip}
\#&Identified features&$z$&$W_\lambda$&B&M&SFR\\
 & & &(\AA)& & &$(M_\odot/yr)$\\
\noalign{\smallskip}
\hline\noalign{\smallskip}
B-C&[O\,{\sc ii}]&0.913&11&22.52\\
A&[O\,{\sc ii}]&0.913&6&23.48\\
D&[O\,{\sc ii}]&0.913&41&23.29&4.6&4\\
14&[O\,{\sc ii}]&0.886&12\\
15&G-band ?&0.223\\
16\_1&[O\,{\sc ii}]&1.268&22&24.21&1.7&4\\
16\_2&[O\,{\sc ii}]&1.082&20&24.13&1.7&3\\
17&[O\,{\sc ii}]&0.859&38&24.04&2.4&3\\
18&G-band, Mg\,{\sc i} ?&0.226& &23.70\\
19& &?& &23.39\\
20&[O\,{\sc ii}], no continuum detected&0.647&$<$100&$>26$\\
21&Mg\,{\sc ii}, [O\,{\sc ii}], [O\,{\sc iii}] 4959\AA\ and 5007\AA\ &0.643&44&2
3.48&2.7&3\\
22&[O\,{\sc ii}]&0.790&17&24.11&2.7&1\\
23&H$\gamma$, [O\,{\sc iii}] 4959\AA\ and 5007\AA\ &0.221& &24.31\\
24\_1&Ca\,{\sc ii} K, Ca\,{\sc ii} H, H$\gamma$&0.631&0\\
24\_2&[O\,{\sc ii}], [O\,{\sc iii}] 5007\AA\ &0.398&29\\
\noalign{\smallskip}
\hline
\end{tabular}
\end{flushleft}
\vskip -1.2cm
\hskip 8.7cm $\Bigl\rbrace$ 23.00$^\ast$

\vskip 0.5cm
$^\ast$ overlapping objects.

\end{table*}

Remarkably, the two slits \#16 and \#24
clearly display complex spectra, with two superimposed objects in 
each. Thanks to the high resolution HST images (Figure 4), it is possible to 
assess each identified redshift to individual objects. In the slit \#16,
the two detected emission lines are spatially slightly separated and we can
clearly identify the most distant of the two objects as the northern one. 
In the slit \#24, it is more difficult because the two objects are well 
superimposed in the direction of dispersion. But the difference in the 
spectral content allows to claim that the reddest galaxy  
is also the most distant one, while the bluest one corresponds 
to an object emitting the strong [O\,{\sc ii}] line. Paradoxically, 
the brightest object is the western galaxy at $z=0.631$, while the faintest 
and most elongated one is the eastern galaxy at $z=0.398$. 

\begin{figure*}
\psfig{figure=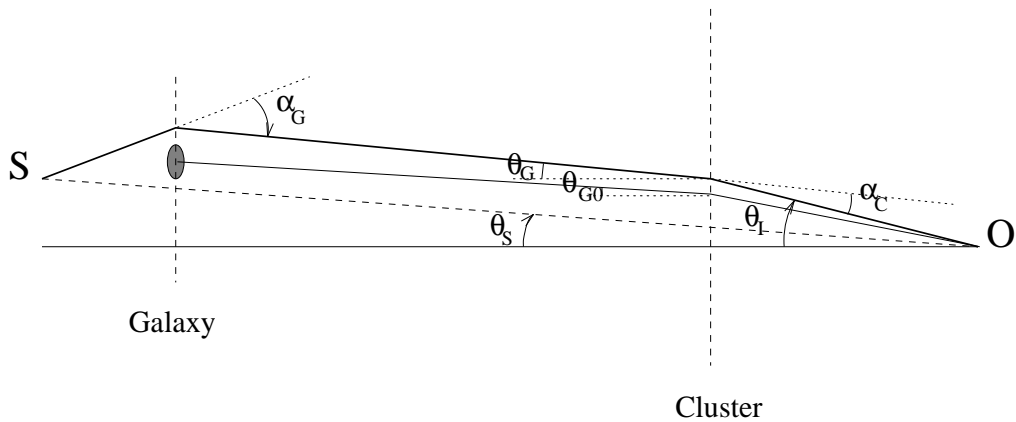,height=8cm}
\caption{Two-screen lensing configuration with a dominant potential in the
cluster plane ($z$=0.231) and an additional one in the background.
}
\end{figure*}

\section{Implications for the lensing model of A2390}
\subsection{Evidence for a complex mass distribution}
From the lensing point of view, it may 
be surprising to find a redshift value of 0.64 for the two arclets \#20
and \#21 which are located behind the giant arc from the cluster center, and
parallel to it. If we consider that the critical line at $z=0.913$ falls
near the giant arc, it is difficult to build a simple lens model with another
critical line at a lower redshift and more distant from the center (see
Fort et al., 1996, in preparation). In
particular arclet \#20 is probably multiple-imaged: a second component
is present on the other side of the neighbouring galaxy (Figure 4) with
similar morphology and color index ($B-R=3.37$ for \#20 and $B-R=3.20$
for the symetric image, Pell\'o et al. 1991), and a possible third image 
is located near this same galaxy, within its enveloppe.
Spectroscopic confirmation is of course required to confirm this assumption, 
which could have strong implications in the lens modelling, as multiple 
images bring strong constraints on the location of the critical curves. 
This could also indicate that the mass distribution is more complex than 
originally suspected, and that the group at $z \simeq 0.64$ may add a 
significant contribution to the lensing of galaxies behind it. It could
even be the optical counterpart of the X--ray clump identified in
Pierre et al. (1996) and superimposed to the main cluster component. Otherwise, 
the redshift identification could be wrong if
the emission line observed is not [O\,{\sc ii}]. This seems quite unlikely
for arclet \#21 as the [O\,{\sc iii}] doublet is also observed at the same 
redshift. For arclet \#20, it is more tentative because only 
one emission line is detected at $\lambda=6137$\AA\ which is clearly visible 
in the 2-D spectrum, but the underlying continuum is too faint to be analysed. 
Future spectroscopy is required to increase the S/N on this arclet and 
on its counterpart candidate.

\subsection{A two screen lensing application}
Another sub-product of this redshift survey is the detection in two
cases of a pair of galaxies with different redshifts nearly aligned on 
the same line of sight, namely the objects in slits \#16 and \#24. 
We propose an attempt to estimate the mass of each of the
less distant one with a simplified analysis of the two-screen
gravitational
lens formalism. We follow the equations of Blandford and Kochanek (1987) 
and the details are given in Appendix.  
By assuming a singular circular isothermal potential (velocity
dispersion
$\sigma$) for the lensing galaxy and using
the modelling of Pierre et al. (1996) for the cluster, we can derive a
relation between $\sigma$, the axial ratios $b_I/a_I$ of the image and
$b_S/a_S$ of the source of the most distant galaxy. 
We plot in Figure 3 the angles $\theta_G$ and $\theta_{G0}$
corresponding to the light travel between the two potentials,
originating
respectively from the lensing galaxy and the background one.
We also call $\varphi$ the angle between the cluster shear axis and the
galaxy shear axis, projected in the sky plane.

In the case of the two galaxies in the slit \#16, the convergence  
and the shear of the cluster projected in the source plane at $z=1.268$
are $\kappa = 0.224$ and $\gamma = 0.151$, 
while $\varphi$ is approximatively 
$49^\circ$ and $\left| \theta_G - \theta_{G0} \right|$ is 0.91\arcsec.
What we want is to propose an estimate of the mass of the perturbing 
galaxy \#16$\_$2 (or equivalently its velocity dispersion $\sigma$) derived 
from the distorsion of the background galaxy \#16$\_$1. 
The main point is that this source at $z=1.268$ appears nearly circular in 
the image plane ($b_I/a_I \simeq 1$), which means that its intrinsic shape is 
elongated in the direction perpendicular to the main axis of magnification. 
We see immediately that without the addition of the galaxy at $z=1.082$, 
the axis ratio of the source is 0.67, a typical value of
0.7 being generally admitted for faint field galaxies (Miralda--Escud\'e,
1991). The addition of the galaxy will even decrease this ratio,
so from a conservative point of 
view we restrict our study to the case where $b_S/a_S$ is larger than 
0.5. This immediately can be translated into a limiting value for the velocity 
dispersion of the lensing galaxy of about 440 km \,s$^{-1}$. This does not give
a strong constraint on the mass of this distant galaxy, essentially 
because the lensing configuration is not quite favorable. Indeed, the 
perturbing galaxy plane is not distant in redshift from the source plane, 
which makes the lens poorly efficient to distort the source. Moreover, the 
cluster effect is rather strong, and already imposes the shape of 
the source being in the low values tail of the axis ratio distribution. 
Nevertheless, this can be also understood in terms of galaxy shape 
evolution as recent results from the Hubble Deep Field seem to show a strong 
increase in 
the ellipticity of faint and distant field galaxies. The conversion between 
$\sigma$ and a total mass is rather tentative, because 
we most probably deal with a spiral galaxy more than with an elliptical
one. But we can derive a crude estimate of this mass from the integration of 
the isothermal potential inside a typical radius of 10 kpc, which is
reasonable for a large spiral galaxy. This gives an upper mass estimate of
$9.0 \hskip 1mm 10^{11} \, M_\odot$ and consequently 
$M/L_V < 29 h$, if we include the galactic
absorption and the k-correction in the absolute V
luminosity measured from the HST image ($V_{555} = 23.59$). 

\begin{figure*}
\psfig{figure=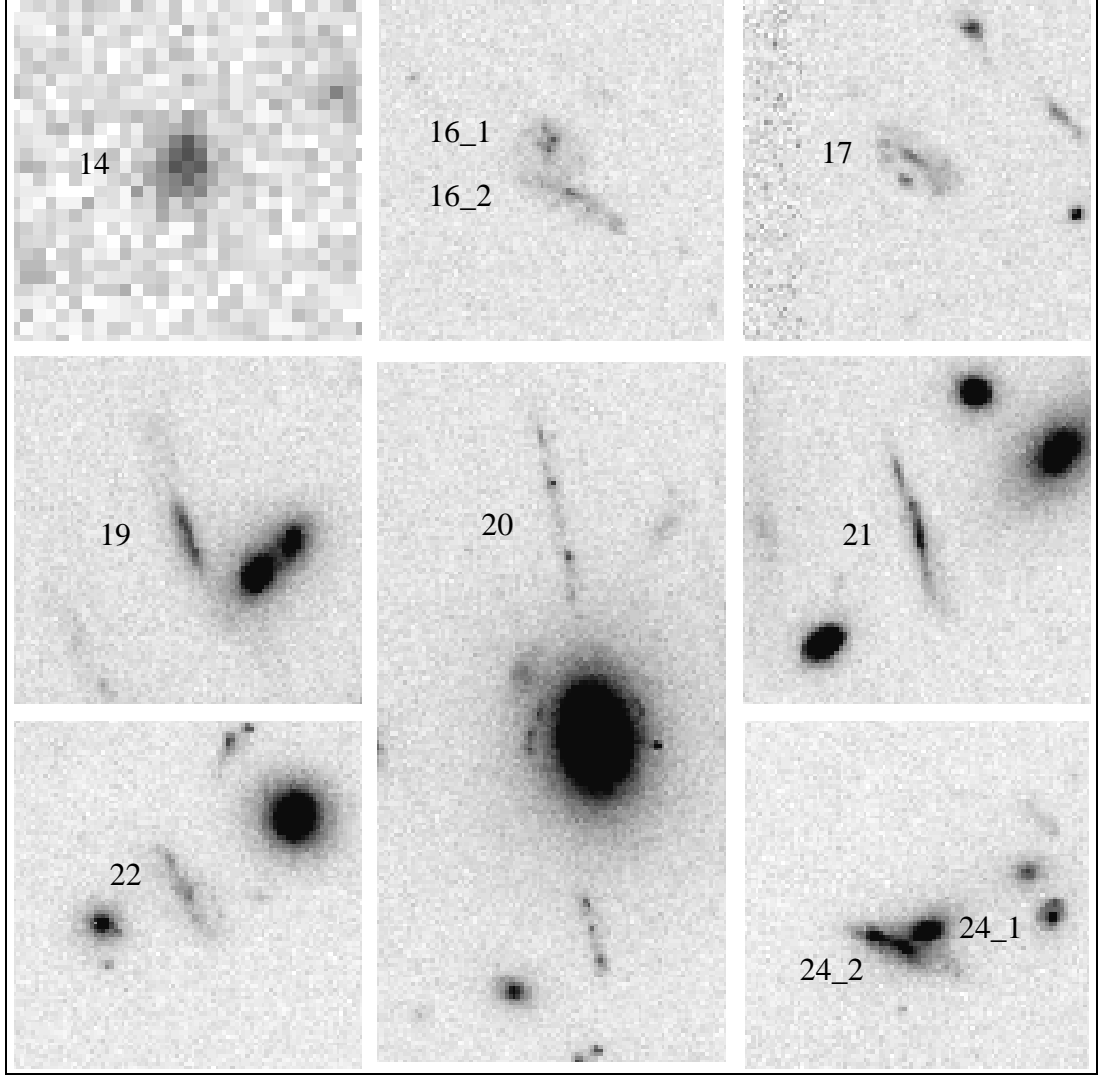,height=15cm}
\caption{Zoom on the morphology of the arclets in A2390, from the HST
deep
images. Each raster is a 8\arcsec$\times$ 8\arcsec\ 
subimage centered
on the arclet. The elongation of some of them is quite spectacular, with
significant bright dots along them, probably characteristics of bright
H\,{\sc ii}
regions 
(see \#20 and \#21). Object \#20 is most probably double imaged,
with another image on the opposite side of the nearby galaxy.
Object \#14 comes from a R-CFHT image, as it lies outside the HST field
of view.}
\end{figure*}

If we apply the same method for the double system of the slit \#24, the 
results are more satisfying in terms of mass estimates: 
numerically, the
convergence of the cluster-lens, computed at $z=0.631$ is 0.246, and the 
shear is 0.105. The angle $\varphi$ 
is $94\degr$, and the distance between the two objects, in the galaxy plane, 
is 0.35\arcsec. This system is also slightly different from the previous 
one in the fact that the image of the background source is rather elongated 
($b_I/a_I = 0.62$), in a direction not far from the main magnification axis.
Without the addition of the galaxy at $z=0.398$, the axis ratio of
the source is 0.82 and  
if we again assume that the axis ratio of the source is higher than 0.5,
the velocity dispersion has to be lower than 140 km \,s$^{-1}$. 
This gives an upper mass estimate of
$9.1 \hskip 1mm 10^{10} \, M_\odot$ and consequently 
$M/L_V < 12 h$, if we include the galactic 
absorption and the k-correction in the absolute V
luminosity measured from the HST image ($V_{555} = 22.8$). 

We insist on the fact that our mass estimates give an upper limit on the
$M/L$. For the system \#16, this limit is not very strong although there
exist only a few examples of the mass estimate of a galaxy at
high redshift. For the system \#24, the lensing configuration is more
favorable to bring a stronger constraint on the $M/L$, which fall within
standard values for spiral galaxies. Determining galaxy mass using weak
gravitational lensing also seems to be quite promising (Brainerd et al.
1995) but the approach is essentially a statistical one, while our 
method applies in a few pecular cases. Moreover, the multiple lenses
formalism we have presented will be useful for a 
future modelling of the cluster if we add the effects of the two mass planes 
$z=0.64$ and $z=0.90$ to analyse the distribution of the very high--z
galaxies behind the cluster.

\section{Star formation rates}
In the spectroscopic content of our sample of distant galaxies, 
we note that among the 12 background objects listed in Table 1, 
10 emit the [O\,{\sc ii}] line characteristic of star forming H\,{\sc ii} 
regions. This high proportion follows the increase in the number of star 
forming galaxies with apparent magnitude shown in Broadhurst et al. (1992). 
An increase of [O\,{\sc ii}] equivalent width with redshift until
$z$=0.4--0.6 has also been reported by Broadhurst et al. (1988) and Colless
et al. (1990). [O\,{\sc ii}] luminous galaxies 
have also been discovered by Cowie et
al. (1995) at redshifts greater than 1.
One way to investigate more quantitatively this relation is to try to 
compute the star formation rate in these lensed-galaxies. Despite the fact 
that H$\alpha$ is the best estimator to infer starbursts intensities 
(Kennicut, 1983), it is not possible to measure this line for redshifts larger 
than 0.4 when spectra are obtained in the optical band only. However, a rough 
estimate can be made with [O\,{\sc ii}] equivalent width for high redshift 
objects using the relation (Kennicut, 1992):
$$SFR\, (M_\odot/yr) = 7 \hskip 1mm 10^{-12} L_{B} W_\lambda$$
where $L_{B}$ is the blue luminosity in solar units on the continuum 
($\lambda_0 \simeq$ 4400\AA) and $W_\lambda$ the rest frame [O\,{\sc ii}] 
equivalent 
width listed in Table 1. In the case of lensed galaxies, one has first to 
correct from the magnification factors taken from the 
gravitational potential proposed in Pierre et al. (1996) for the central 
part of the cluster. Rest frame B luminosities were determined with B 
magnitudes measured on an image taken at CFHT on November 1991, then 
k-corrected with a template Sd spectrum (Bruzual and Charlot, 1993) except 
for object D corresponding to a Sc galaxy (Pell\'o et al. 1991). 
Interstellar extinction was taken into account ($A_{B}$=0.28 for A2390, 
Burstein and Heiles, 1982). 

The difficulty is to estimate the uncertainties on the SFR which 
could be large for many reasons.
First the extinction by the galaxy itself is unknown, an average value of 1
magnitude was adopted by Kennicut (1992) for the extinction of H$\alpha$. 
This can lead to important differences for individual objects
but the SFR averaged over all arclets should not strongly suffer this bias. 
Second, there is an intrinsic dispersion in the relation between [O\,{\sc ii}] 
and H$\alpha$ line fluxes which do not represent exactly the same quantities, 
as raised originally by Kennicut (1992). 
Additional uncertain parameters are the exact proportion of
ionizing photons escaping from the H\,{\sc ii} regions, and the slope of the 
IMF. Globally, this can amount to an error of nearly 100\%\ or a factor 
of two. 
Finally, our corrections for the absolute luminosity depends strongly 
on the gravitational magnification, which may contain a source of 
uncertainty as far as an overconstrained model has not emerged. The 
results of this evaluation are also listed in Table 1. 

The mean star formation rate among our sample of distant galaxies 
is $3\, M_\odot/yr$, with a factor 2 of uncertainty, a value of the same order 
of magnitude as for nearby blue galaxies (Kennicut, 1983). Gallagher et al. 
(1989) also measured [O\,{\sc ii}] equivalent widths in nearby galaxies but 
their derived SFR are lower. 
In the case of more distant sources, 
a tentative evaluation of the SFR was proposed by Mellier et al. (1991)
for the very distant source of the giant arc in Cl2244--02 at $z=2.24$, with 
a value varying from 7 to 20 $M_\odot/yr$ depending on the correction factors. 
Note that this value was obtained from the measurement of the UV continuum, 
with a slightly different calibration procedure. 
A more detailed evaluation has been recently proposed
by Ebbels et al. (1996), from a complete spectral synthesis applied on one 
arclet in A2218 identified at $z=2.515$, giving a SFR of about 10 $M_\odot/yr$. 
Steidel et al. (1996) also found similar values based on the far--UV
continuum in objects at redshifts of 3--3.5.
In any case, starburst activity in high redshift galaxies seems to be more
frequent than in the local universe but its intensity is comparable to
nearby spiral galaxies, at least up to $z\sim 1$ with maybe a global increase 
at higher $z$. It remains much below the values 
found in several objects at $z=1.8-2.0$ such as radio galaxies or damped 
Ly$\alpha$ absorbers which display hundreds of $M_\odot/yr$ of SFR 
(Djorgovski, 1988; Elston et al., 1991). But these objects belong to other 
classes of distant galaxies with signs of strong activity and evolution 
and they cannot be used as a direct probe for the evolution of the star 
formation activity of field galaxies. 

\section{Conclusions}
In this paper, we have presented 9 new redshifts of gravitationally 
lensed background galaxies located in the same field, behind the cluster-lens 
A2390, with a success rate of 90\%\ for the redshift determination of these 
faint objects. We suspect the existence of two peculiar planes 
of background objects, corresponding to spatial structures at $z=0.64$ and 
0.90. But a better statistics is required to confirm them and to 
quantify their mass distribution and their influence on the lens configuration. 
We also confirm the general trend that the population of arclets is  
mostly distributed in the range 0.6$<z<$1 and their starburst
activity is more frequent than in nearby galaxies, although not significantly 
stronger. 
Constraints on $M/L$ for two galaxies at $z=0.398$ and $z=1.082$ have
been obtained thanks to multiple light deflection by two gravitational
potentials at different redshifts.  
The next step is now to introduce these confirmed arclets as constraints on 
an updated lens model of the cluster potential obtained together with the 
analysis of the HST images of the arcs and arclets. The spectroscopic results 
will give constraints mainly on the slope of the mass profile outside the 
critical radius. Combined with a new modelling, the {\sl lensing redshift} 
method will then be able to determine for a set of new arclets the most 
probable redshifts that can account for their location and shape. Used as 
a bootstrap technique, it may give the unique 
opportunity to get the redshift of galaxies 3 magnitudes fainter than the 
faintest redshift surveys. Similar results are also beeing tested on the 
cluster A2218 which displays a similarly large number of arclets (Kneib
et al. 1996). This method gives a statistical redshift distribution. The 
combination of the two lenses and possibly of a few other ones such as
A370 and AC114 will greatly increase the sample of faint 
galaxies. One of their main interest is the small selection biases introduced 
in their identification (no color selection) and the window they open to the high redshift 
tail of the faint field galaxy population.

Gravitational lensing has brought a large number of results on the mass 
distribution of clusters of galaxies. Now the window is open to look at 
the population of the sources and to benefit from their magnification to 
have access on new informations on their redshift distribution and on their
evolution both in spectral content and morphology.

\acknowledgements 
We wish to thank Y. Mellier for his strong encouragements and remarks during 
the writting of the paper. We are also very grateful to
J.F. Le Borgne and R. Pell\'o for fruitful discussions 
about A2390 and to J.P. Kneib for useful comments on
the manuscript, especially on the double lens equations and formalism. This 
work was partly supported by the Groupe de Recherche ``Cosmologie''.

\vskip 0.5cm
{\bf \hskip -5mm Appendix}
\vskip 0.2cm

The lens equation in a system with
two redshift planes can be written as (Blandford and Kochanek, 1987):
\begin{equation}
\vec{\theta_S} = \vec{\theta_I} - {D_{CS}\over D_{OS}} \vec{\alpha_C}
(\vec{\theta_I}) - {D_{GS}\over D_{OS}} \vec{\alpha_G} (\vec{\theta_G} -
\vec{\theta}_{G0})
\end{equation}
with the angles defined in Figure 3.
$\vec{\theta_G} - \vec{\theta}_{G0}$ is the angle between the light paths
originating from the lensing galaxy and the background one in the
region between the two potentials.
The Hessian of the transformation, which is the inverse of the
amplification matrix is:
 \begin{equation}
{\cal H} = {\cal I} - {D_{CS}\over D_{OS}} \, {\partial \vec{\alpha_C}
\over \partial \vec{\theta_I}} - {D_{GS}\over D_{OS}} \, {\partial
\vec{\alpha_G} \over \partial \vec{\theta_G}} \, {\partial
\vec{\theta_G}
\over \partial \vec{\theta_I}}
   \end{equation}
The first two terms of this relation can be related to the magnification
matrix of the cluster ${\cal H}_C$, which is written, when projected on the
local magnification axis:
\begin{equation}
{\cal H}_C = {\cal I} - {D_{CS}\over D_{OS}} \, {\partial \vec{\alpha_C}
\over \partial \vec{\theta_I}}
           = \left( \begin{array}{cc} 1-\kappa - \gamma & 0 \\
                                      0 & 1-\kappa + \gamma
                    \end{array} \right)
   \end{equation}
$\kappa$ is the convergence of the lens
and $\gamma$ is the local shear computed at the source redshift with
the modelling of Pierre et al. (1996).
For the perturbative lens which corresponds to an individual galaxy, we
associate a singular isothermal sphere potential with a velocity
dispersion
$\sigma$.
In a polar coordinate system where axis 1 is the tangential axis and
axis 2 is
the radial axis,
the magnification matrix takes a simple form (Kneib 1993):
 \begin{equation}
{\cal H}_G = {\cal I} - {D_{GS}\over D_{OS}} \, {\partial \vec{\alpha_G}
\over \partial \vec{\theta_G}}
           = \left( \begin{array}{cc} 1- 4\pi \, {D_{GS}\over D_{OS}} \,
          {\sigma^2 \over c^2} \, {1 \over \left| \theta_G -
            \theta_{G0} \right|} & 0 \\
            0 & 1   \end{array} \right)
\end{equation}
We now want to combine these two lenses, and have to project the matrix
${\cal H}_G$ on the main axis of ${\cal H}_C$ before any addition. Let
$\varphi$ be the angle between the cluster shear and the galaxy shear.
Moreover,
as seen in Figure 3, $\vec{\theta_G}$ and $\vec{\theta_I}$ are related
such that
$\vec{\theta_G} = \vec{\theta_I} - \vec{\alpha_C}$ in the cluster plane.
Consequently, we have:
\begin{equation}
{\partial \vec{\theta_G} \over \partial \vec{\theta_I}} = {\cal I} +
\left(
{\cal H}_C - {\cal I} \right) \, {D_{OS}\over D_{CS}} \equiv \cal M
\end{equation}
and finally,
 \begin{equation}
{\cal H} = {\cal H}_C + {\cal R}ot (-\varphi) \, \left( {\cal H}_G -
{\cal I} \right) {\cal R}ot (\varphi) \, {\cal M}
\end{equation}
which can be expressed as 
\begin{equation}
{\cal H} = \left( \begin{array}{cc} h_{11} & h_{12} \\
h_{21} & h_{22}
\end{array} \right)
\end{equation}
with
\begin{equation}
h_{11}=1- \kappa -\gamma - C_+ \sigma^2\cos^2 \varphi
\end{equation}
\begin{equation}
h_{12}={1 \over 2} C_- \sigma^2 \sin 2\varphi
   \end{equation}
\begin{equation}
h_{21}={1 \over 2} C_+ \sigma^2 \sin 2\varphi
\end{equation}
\begin{equation}
h_{22}=1 - \kappa +\gamma - C_- \sigma^2 \sin^2 \varphi
   \end{equation}
and
\begin{equation}
C_{\pm} = {4\pi \over c^2} \, {D_{GS}\over D_{OS}} \,{1 \over \left|
\theta_G -
\theta_{G0} \right|} \left( 1 - \left(\kappa \pm \gamma \right) {D_{OS}
\over D_{CS}} \right)
\end{equation}
As pointed out by Kochanek and Apostolakis (1988), lensing by several
redshift
planes makes the magnification matrix asymetric, so the formalism of a
single screen has to be used with care. We propose some estimates of the
semi--major and semi--minor axis of the source using essentially the
shape matrix formalism for
the source and the image (Kochanek, 1990), which are related by:
\begin{equation}
{\cal F}_S = {\cal H} \, {\cal F}_I \, ^t{\cal H}
\end{equation}
In the two cases we analyse here, the orientation of the
image falls near the main axis of the shear, so that ${\cal F}_I$ is
diagonal with simple terms $a_I^2$ and $b_I^2$. So whatever the true
orientation of the source in the source plane, we can write:
 \begin{equation}
{\rm det} {\cal F}_S = a_S^2 \, b_S^2 \qquad {\rm and} \qquad
{\rm Tr} {\cal F}_S = a_S^2 + b_S^2
\end{equation}
This can be easily solved for $b_S/a_S$, knowing $b_I/a_I$, for each
value of $\sigma$.

\end{document}